\documentclass[aps, pra, twocolumn, superscriptaddress]{revtex4-1}
\usepackage{graphicx}
\usepackage{amsmath}
\usepackage{amssymb}
\usepackage{bm}

\usepackage[colorlinks, linkcolor=blue, urlcolor=magenta, citecolor=blue]{hyperref}

\begin{document}

\title{Enhanced Hamiltonian Learning Precision with Multi-Stage Neural Networks}
%\title{Multi-Stage Neural Networks for Enhanced Precision in Hamiltonian Learning}

\author{Zhengjie Kang}
\affiliation{Guangdong Provincial Key Laboratory of Quantum Metrology and Sensing \& School of Physics and Astronomy, Sun Yat-Sen University (Zhuhai Campus), Zhuhai 519082, China.}

\author{Hao Li}
\affiliation{Guangdong Provincial Key Laboratory of Quantum Metrology and Sensing \& School of Physics and Astronomy, Sun Yat-Sen University (Zhuhai Campus), Zhuhai 519082, China.}

\author{Shuo Wang}
\affiliation{Guangdong Provincial Key Laboratory of Quantum Metrology and Sensing \& School of Physics and Astronomy, Sun Yat-Sen University (Zhuhai Campus), Zhuhai 519082, China.}

\author{Jiaojiao Li}
\affiliation{Guangdong Provincial Key Laboratory of Quantum Metrology and Sensing \& School of Physics and Astronomy, Sun Yat-Sen University (Zhuhai Campus), Zhuhai 519082, China.}

\author{Yuanjie Zhang}
\affiliation{Guangdong Provincial Key Laboratory of Quantum Metrology and Sensing \& School of Physics and Astronomy, Sun Yat-Sen University (Zhuhai Campus), Zhuhai 519082, China.}

\author{Zhihuang Luo}
\email{luozhih5@mail.sysu.edu.cn}
\affiliation{Guangdong Provincial Key Laboratory of Quantum Metrology and Sensing \& School of Physics and Astronomy, Sun Yat-Sen University (Zhuhai Campus), Zhuhai 519082, China.}

%\date{\today}

\begin{abstract}
Learning quantum Hamiltonians with high precision is important for quantum physics and quantum information science. We propose a multi-stage neural network framework that significantly enhances Hamiltonian learning precision through successive network optimization of residual errors. Our approach utilizes time-series data from single-qubit Pauli measurements of random initial states, enabling the estimation of unknown Hamiltonian parameters without prior structural assumptions. We demonstrate the framework on two-qubit systems, achieving orders-of-magnitude improvement in parameter accuracy, and further extend the method to larger systems by integrating dynamical decoupling techniques. Additionally, the protocol exhibits robustness against experimental noise. This work bridges the gap between scalable Hamiltonian learning and high-precision requirements, offering a practical tool for precise quantum control and metrology.
\end{abstract}

\maketitle

\section{Introduction}

Precise identification of quantum Hamiltonians from their dynamics is a fundamental task for quantum physics, with extensive applications in quantum computing~\cite{nielsen2010quantum, tilly_variational_2022, Cerezo2022, fisher_random_2023}, quantum simulation~\cite{Lloyd1996, RevModPhys.86.153, Daley2022, Buluta2009}, quantum materials~\cite{Keimer2017}, and quantum metrology~\cite{Leibfried2004, PhysRevLett.88.231102, PhysRevLett.71.1355, PhysRevA.54.R4649}. 
Previous methods for Hamiltonian learning, such as quantum process tomography~\cite{PhysRevA.77.032322, PhysRevA.64.012314, li_hamiltonian_2020, chen_experimental_2021, altepeter_ancilla-assisted_2003}, face scalability challenges due to their exponential resource demands. To tackle these challenges, various alternative methods have been proposed. For instance, several studies advocate for learning Hamiltonian coefficients by solving linear equations \cite{bairey_learning_2019, PhysRevLett.113.080401, wang_quantum_2018}, where the coefficient matrix is derived from local measurement outcomes. However, the effectiveness of these approaches is significantly influenced by the spectral gap of the coefficient matrix, which is often poorly characterized. While compressed sensing \cite{rudinger_compressed_2015, shabani_estimation_2011} and dynamical decoupling (DD) \cite{burgarth_coupling_2009, gu_practical_2024, huang_learning_2023} mitigate these limitations by exploiting sparsity or partitioning interactions, their efficacy hinges on prior assumptions about system structure or spectral properties. Quantum simulators combined with Bayesian inference offer alternative pathways but remain constrained by hardware noise and computational overhead \cite{wang_experimental_2017, wiebe_hamiltonian_2014}.

Recent advances in machine learning (ML) have introduced powerful tools for quantum system characterization \cite{krenn_artificial_2023, wang_predicting_2024, PhysRevResearch.4.013097, harney_entanglement_2020, PhysRevA.103.032406, PhysRevLett.121.167204, PhysRevLett.114.200501, PhysRevLett.121.255304, cimini_neural_2020, glasser_neural-network_2018}. Neural networks have demonstrated success in tasks ranging from quantum state tomography \cite{ahmed_quantum_2021, torlai_neural-network_2018, carleo_solving_2017, PhysRevA.98.012315, wang_hybrid_2023, kieferova_tomography_2017} to quantum phase transition \cite{scheurer_unsupervised_2020, rodriguez-nieva_identifying_2019, PhysRevB.94.195105, PhysRevB.97.134109, vanNieuwenburg2017} and topological phase classification \cite{PhysRevLett.120.066401, lian_machine_2019, zhang_machine_2021, carrasquilla_machine_2017}. Notably, ML models have been adapted to directly infer Hamiltonian parameters from time-resolved measurements \cite{che_learning_2021, kwon_magnetic_2020, bienias_meta_2021, shi_parameterized_2023, wilde_scalably_2022, cao_supervised_2020, han_tomography_2021}, circumventing the need for exhaustive tomography. However, existing approaches often struggle with scalability to multi-qubit systems, sensitivity to parameter magnitude disparities, and reduction of ML errors. The single neural networks used for training readily reach the precision limits due to the local minima, even with large network size and extensive training iterations  \cite{baldi_neural_1989, wang_multi-stage_2024}. 

In this work, we propose a multi-stage Hamiltonian learning framework to achieve higher-precision estimation of parameters. The neural networks for Hamiltonian learning involve multiple stages and are 
successively optimized by the residues from the preceding stage. For training the network in each stage, we start with random initial states and obtain time-series data by measuring the single-qubit Pauli operators. The unknown parameters underlying the dynamical evolutions of the Hamiltonians can be estimated from these single-qubit measurements. We demonstrate our multi-stage Hamiltonian learning framework for two-qubit systems and then apply these pre-trained models to estimate the parameters of larger Hamiltonians by integrating the DD techniques. We further analyze the robustness against the noises. 
Our protocol enhances the Hamiltonian learning precision with several orders of magnitude and has practical scaling for large systems.

\section{Multi-stage Hamiltonian learning framework}

\begin{figure*}
    \centering
    \includegraphics[width=0.9\textwidth]{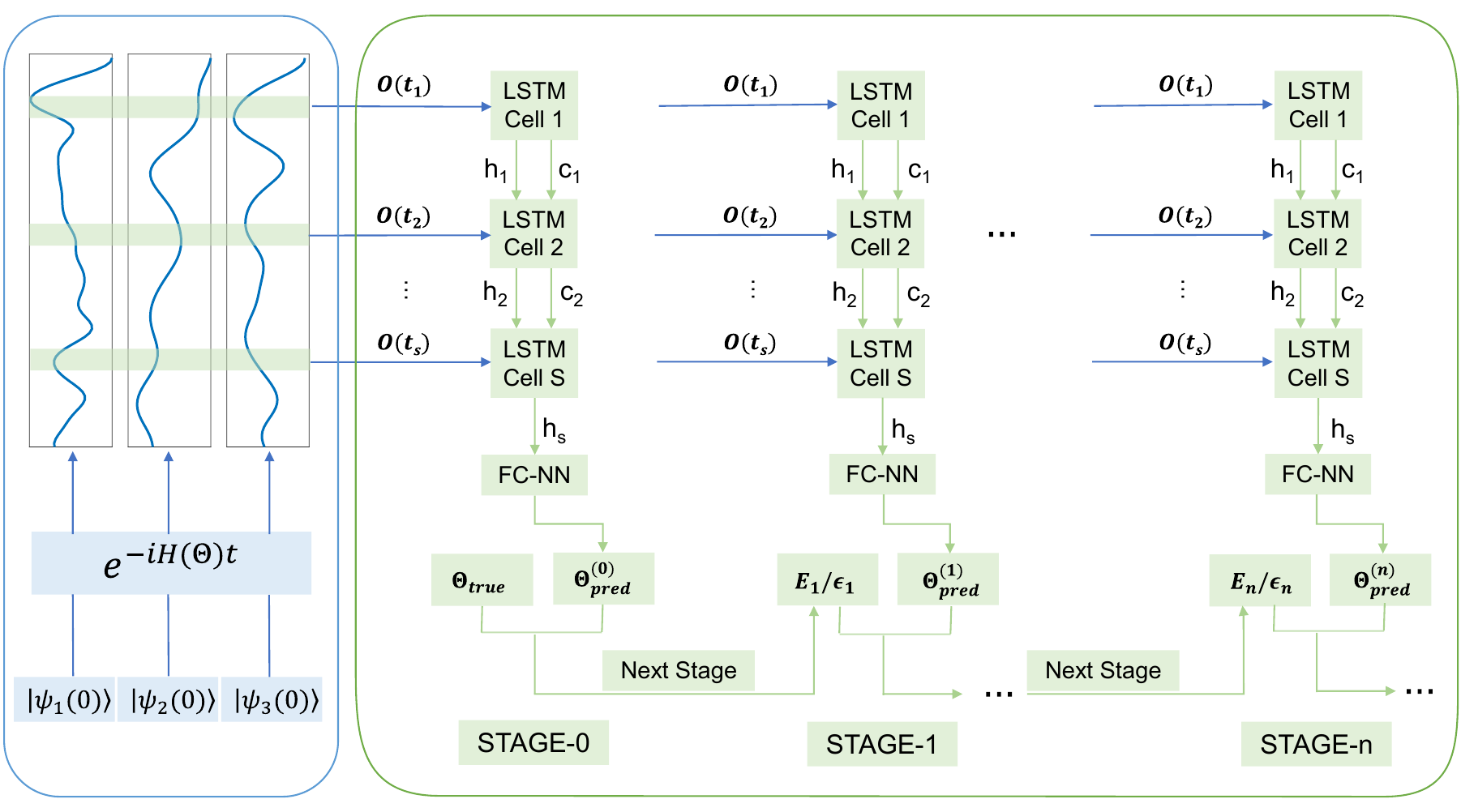} 
    \caption{Schematic diagram of the multi-stage Hamiltonian learning framework. The system begins with three random initial states $|\psi_i(0)\rangle$ for $i=1, 2, 3$, and freely evolves under an unknown Hamiltonian $H(\bm{\Theta})$. The resulting time series observations $\bm{O}(t)$ obtained by measuring the single-qubit operators are input to the LSTM cells of different stage layers. The FC-NNs output the predicted values $\bm{\Theta}_{\text{pred}}^{(n)}$ ($n=0, 1, \cdots, n$) for target parameters $\bm{\Theta}_{\text{true}}, \bm{E}_1/\epsilon_1, \cdots, \bm{E}_n/\epsilon_n$. Here $\bm{E}_n$  represents the error between the predicted and target parameters, and $\epsilon_n=\|\bm{E}_n\|/\sqrt{M}$ denotes the corresponding normalization factor.}
    \label{fig: mshl}
\end{figure*}

To set the stage, we first give a formal definition of the Hamiltonian learning problem.
We consider a Hamiltonian on an $N$-qubit system
\begin{equation}\label{eq: Ham}
H(\bm{\Theta}) = \sum_{i=1}^M \theta_i T_i,
\end{equation}
with an unknown parameter vector $\bm{\Theta} \equiv [\theta_1,\cdots, \theta_M]^T$, and  $T_i \in \{I, \sigma_x, \sigma_y, \sigma_z\}^{\otimes N}$ denotes the tensor product of Pauli operators. The Hamiltonian is assumed to be traceless (i.e., $T_i \neq I^{\otimes N}$), and the structure $\{T_i\}$ of the Hamiltonian is known. The Hamiltonian learning problem aims to infer all unknown coefficients $\theta_i$s through time-series local observables under the dynamical evolution of the Hamiltonian (\ref{eq: Ham}). 

The way to solve this problem is as follows. Starting with an initial pure state $\rho(0)=|\psi_i(0)\rangle\langle\psi_i(0)|$, the system undergoes free evolution under the unknown Hamiltonian (\ref{eq: Ham}). The state at $t$ is $\rho(t) = e^{-iH(\bm{\Theta})t}\rho(0)e^{iH(\bm{\Theta})t}$. By measuring single-qubit operators $\sigma_{\alpha}^j$, we obtain a series of observations:
\begin{equation}
O_{\alpha}^j(t) = \text{Tr}(\sigma_{\alpha}^j\rho(t)) = \text{Tr}(\sigma_{\alpha}^j(t)\rho(0)),
\end{equation}
for $\alpha = x, y, z$ and $j= 1, \cdots, N$. Here we use the cyclic property of the trace, and
%\begin{equation}
	$\sigma_{\alpha}^j(t) \equiv e^{iH(\bm{\Theta})t}\sigma_{\alpha}^j e^{-iH(\bm{\Theta})t} = \sum_{m=0}^{\infty}\frac{(it)^m}{m!} \text{ad}_{H(\bm{\Theta})}^m(\sigma_{\alpha}^j)$,
%\end{equation}
where $\text{ad}_{H(\bm{\Theta})}(\cdot)=[H(\bm{\Theta}), \cdot]=\sum_{i=1}^M \theta_i [T_i, \cdot]$ represents the adjoint operator.
Thus, the single-qubit observations can be expressed as a power series in time, i.e.,
\begin{equation}
 O_{\alpha}^j(t) = O_{\alpha}^j(0) +\sum_{m=1}^\infty c_m t^m,
\end{equation}
with $c_m = \frac{i^m}{m!}\sum_{i_1=1}^M\cdots\sum_{i_m=1}^M a_{i_1,\cdots, i_m}\theta_{i_1}\cdots\theta_{i_m}$, and $a_{i_1, \cdots, i_m} = \text{Tr}\{[T_{i_1},\cdots, [T_{i_m}, \sigma_{\alpha}^j] \rho(0)\}$.
This indicates that the Hamiltonian parameter $\theta_i$ will contribute to the dynamical evolution of the system if $c_m \neq 0$. By analyzing the time-series observation data from the single-qubit measurements, we can infer all unknown Hamiltonian parameters $\bm{\Theta}$ using neural networks. 
To achieve this, we choose three random initial states $|\psi_i(0)\rangle$ for $i=1,2,3$, ensuring that $a_{i_1, \cdots, i_m}\neq 0$ as much as possible.

The resulting single-qubit time-series observations under the dynamical evolution $e^{-iH(\bm{\Theta})t}$ are collected as a vector
\begin{equation}
   \bm{O}(t) = \{O_{\alpha}^j(t)|\alpha=x, y, z,1\leq j \leq N\},
\end{equation}
which serves as the input data to the neural network of Fig. \ref{fig: mshl}. This network consists of $S$ LSTM cells and a fully connected neural network (FC-NN). Here the total time $T$ is divided into $S=T/\tau$ segments, each with a duration of $\tau$.
The $j$th LSTM cell receives the observation $\bm{O}(t_j)$ at time $t_j$, along with the memory cell $h_{j-1}$ and hidden cell $c_{j-1}$ from the previous time step. After processing, the updated memory cell $h_j$ and hidden cell $c_j$ are passed to the next LSTM cell. This mechanism allows the LSTM to capture temporal dependencies effectively and improve the neural network performance. Finally, the FC-NN layer outputs the predicted result $\bm{\Theta}_{\text{pred}}$ for the current stage. Most existing works stop at this stage, achieving good results regardless of the input data choice. 
However, the precision of $\bm{\Theta}_{\text{pred}}$ can be further enhanced significantly through a multi-stage Hamiltonian learning framework, as demonstrated in Fig.~\ref{fig: mshl}. 
During multi-stage training, a critical issue arises: if the initial output data are on the order of $O(1)$, the error $\bm{E}_n$ decreases progressively, often reaching scales much smaller than $1$. Standard weight initialization methods, such as Xavier initialization, struggle to capture features of data that vary significantly in scale. Therefore, it is essential to normalize the error $\bm{E}_n$ using the root mean square value $\epsilon_n$ defined below.

Figure \ref{fig: mshl} shows the core idea of our multi-stage neural network.
In STAGE-0, we begin with the relationship of the input and output layers
\begin{equation}
    \{\bm{O}(t), \bm{\Theta}_{\text{true}}\}\rightarrow\bm{\Theta}_{\text{pred}}^{(0)},
\end{equation}
which produces the error $\bm{E}_1=\bm{\Theta}_{\text{true}} - \bm{\Theta}_{\text{pred}}^{(0)}$ and the normalization factor $\epsilon_1 = \|\bm{E}_1\|/\sqrt{M}$. We then proceed to the next stage (STAGE-1), 
\begin{equation}
	\{\bm{O}(t), \bm{E}_1/\epsilon_1\}\rightarrow\bm{\Theta}_{\text{pred}}^{(1)}
\end{equation}	
resulting in the new error $\bm{E}_2=\bm{\Theta}_{\text{true}} - \bm{\Theta}_{\text{pred}}^{(0)}-\epsilon_1\bm{\Theta}_{\text{pred}}^{(1)}$ and the updated normalizatioin factor $\epsilon_2 = \|\bm{E}_2\|/\sqrt{M}$.
In general, we define STAGE-$n$ as
\begin{equation}
	\{\bm{O}(t), \bm{E}_n/\epsilon_n\}\rightarrow\bm{\Theta}_{\text{pred}}^{(n)},
\end{equation}
where $ \bm{E}_n = \bm{\Theta}_{\text{true}}-\bm{\Theta}^{(n)}$ and $\epsilon_n = \|\bm{E}_n\|/\sqrt{M}$.
Here 
\begin{equation}
	\bm{\Theta}^{(n)} = \sum_{j=0}^{n-1}\epsilon_j\bm{\Theta}_{\text{pred}}^{(j)} 
\end{equation}
is the final output of predicted Hamiltonian parameters after STAGE-$n$ and $\epsilon_0 \equiv 1$.
This training process is repeated iteratively until no further improvement can be achieved.

The optimization of the multi-stage neural network is achieved by adjusting hyperparameters (weights and biases) through self-feedback to minimize the loss functions at each stage. The loss functions are defined as the mean squared error (MSE) between the true values of the target parameters and the predicted values:
\begin{equation}
L_0 = \|\bm{\Theta}_{\text{pred}}^{(0)} - \bm{\Theta}_{\text{true}}\|^2/M,
\end{equation} 
for STAGE-0, and 
\begin{equation}
	L_{n}= \|\bm{\Theta}_{\text{pred}}^{(n)} - \bm{E}_n/\epsilon_n\|^2/M,
\end{equation}
for STAGE-$n$ ($n\neq 0$).
We employ the Adam algorithm for gradient descent optimization \cite{kingma2014adam, ruder_overview_2017}. Adam is an adaptive learning rate optimization method that leverages the benefits of both Momentum and RMSProp. Unlike traditional Stochastic Gradient Descent, Adam dynamically adjusts the learning rate by estimating both the first-order moment (momentum) and the second-order moment (variance) of the gradient. This adaptive adjustment mechanism smooths the optimization path and mitigates unnecessary oscillations. Adam's efficiency and stability make it the preferred algorithm for tackling complex optimization problems, particularly in non-convex scenarios.

The fidelity of the $n$th-stage Hamiltonian learning is evaluated by calculating the cosine similarity between the predicted parameters $\bm{\Theta}^{(n)}$ and the true values $\bm{\Theta}_\text{true}$, defined as follows:
\begin{equation}\label{eq: fidelity}
F(\bm{\Theta}^{(n)}, \bm{\Theta}_{\text{true}}) = \frac{\bm{\Theta}^{(n)} \cdot \bm{\Theta}_\text{true}}{\|\bm{\Theta}^{(n)}\| \cdot \|\bm{\Theta}_\text{true}\|}.
\end{equation}
Using this optimization approach, our multi-stage NNs can theoretically achieve the numerical precision limits of double-floating-point computation. This has been verified with a one-qubit Hamiltonian that has a single parameter. However, it requires more computational resources including the training samples and the number of neurons, as the complexity of the Hamiltonian model increases.

\section{Examples of two-qubit systems}

As demonstrations of our multi-stage Hamiltonian learning framework, we first consider a two-qubit system with three unknown Hamiltonian parameters. The Hamiltonian can be expressed as follows
\begin{equation}\label{eq: Ham2qb1}
H_1 = \omega_1 \sigma_z^1 + \omega_2 \sigma_z^2 + J_{12} \sigma_x^1 \sigma_x^2,
\end{equation}
where $\omega_1$ and $\omega_2$ represent the frequencies of the external magnetic field applied to qubits $1$ and $2$, respectively, along the $z$-axis. The $J_{12}$ denotes the coupling strength between two qubits along the $x$-direction. Without loss of generality, suppose that $\omega_1, \omega_2\in [-1, 1]$, and $J_{12} \in [-1, 1]$.
We further consider a two-qubit Hamiltonian with nine coupling parameters described as
\begin{equation}\label{eq: Ham2qb2}
 H_2 = \sum_{\alpha, \beta \in \{x, y, z\}} J_{\alpha\beta} \sigma_{\alpha}^1 \sigma_{\beta}^{2},
\end{equation}
where \(J_{\alpha\beta} \in [-1,1]\) represents the coupling strength. This model is highly versatile as it encompasses all two-body interactions.
To complete the multi-stage training, we generated 300,000 datasets, allocating 80\% for the training set and 20\% for the validation set.
For each dataset, the single-qubit observation $\bm{O}(t)$ was discretized into $S=100$ points, with a measurement interval set to $\tau = 0.02\pi$ (see details in Appendix. \ref{seca: Optimization of sampling parameters}). 

\begin{figure}
    \centering
    \includegraphics[width=0.48\textwidth]{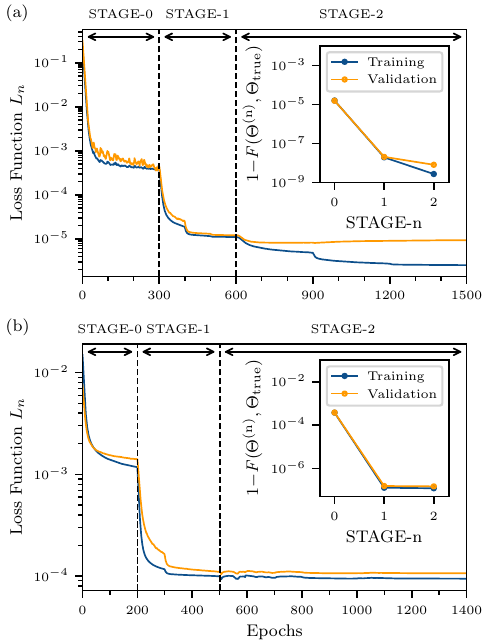} 
    \caption{Results of multi-stage Hamiltonian learning for two-qubit systems with Hamiltonians $H_1$ (a) and $H_2$ (b). The blue and orange lines represent the loss functions $L_n$ ($n=0, 1, 2$) for training and validation across STAGE-0, STAGE-1, and STAGE-2, plotted against epochs. Insets display the infidelities, expressed as $1-F(\bm{\Theta}^{(n)}, \bm{\Theta}_{\text{true}})$, highlighting the dissimilarities between the true Hamiltonian parameters $\bm{\Theta}_{\text{true}}$ and the predicted values $\bm{\Theta}^{(n)}$ after training at each STAGE-$n$. }
    \label{fig: lossfunction}
\end{figure}

Figure~\ref{fig: lossfunction} shows the learning results of two examples using multi-stage neural networks.
In STAGE-0 of Fig.~\ref{fig: lossfunction}(a), the loss function $L_0$ for both the training and validation sets decreases rapidly and stabilizes below $10^{-3}$. Simultaneously, the fidelity increases to over 99.99\%, indicating that the model has successfully captured the primary structure of the target Hamiltonian. However, the single-stage model reaches saturation at this point, with no further significant reduction in loss or improvement in fidelity. 
In STAGE-1, the loss function $L_1$ further decreases to the order of $10^{-5}$, and the fidelity improves to 99.99999\%. Importantly, the performance on the validation set remains consistent with that of the training set, demonstrating the model's strong generalization capability during error learning. 
In STAGE-2, while the loss function $L_2$ for the training set eventually drops to around $10^{-6}$ and the training fidelity approaches an impressive 99.999999\%, the improvement in validation fidelity becomes less significant, as shown in Fig.~\ref{fig: lossfunction}(a). In Fig~\ref{fig: lossfunction}(b), we similarly observe that the training and validation fidelities for STAGE-1 increase significantly from 99.9\% in STAGE-0 to over 99.9999\%, but further gains in STAGE-2 are marginal.
This suggests that the capacity to learn from errors saturates as the stages progress, limiting further optimization. Additionally, computational resource constraints restrict the dataset size, impeding further improvement in validation performance. 
Therefore, the multi-stage Hamiltonian learning progressively refines its predictions by learning residual errors (see Appendix. \ref{seca: Error distributions}), significantly improving training accuracy and fidelity while maintaining excellent generalization. The insets of Fig.~\ref{fig: lossfunction} provide a more intuitive view of the training performance at each stage, validating the effectiveness of our proposed multi-stage training strategy.

For the issues that further improvements are limited in the training of STAGE-2, we can gain deeper insights through correlation analysis (see Appendix. \ref{seca: Data correlations}). For the training data at each stage, we calculate the Pearson Correlation Coefficient (PCC) and Mutual Information (MI) between the input observation data and the target output.
As the training progresses to STAGE-2, the maximum values of PCCs and MIs gradually decrease, approaching zero. This indicates that the correlation between the input data and target output significantly weakens in STAGE-2, making it difficult for the model to extract useful learning signals from the residual error information. This phenomenon explains why increasing the model complexity does not lead to further improvement in STAGE-2 training performance. This finding provides important insights for future improvements in training data generation strategies or optimization of the model architecture.

\section{Applications to larger systems}

\begin{figure*}
    \centering
    \includegraphics[width=0.95\linewidth]{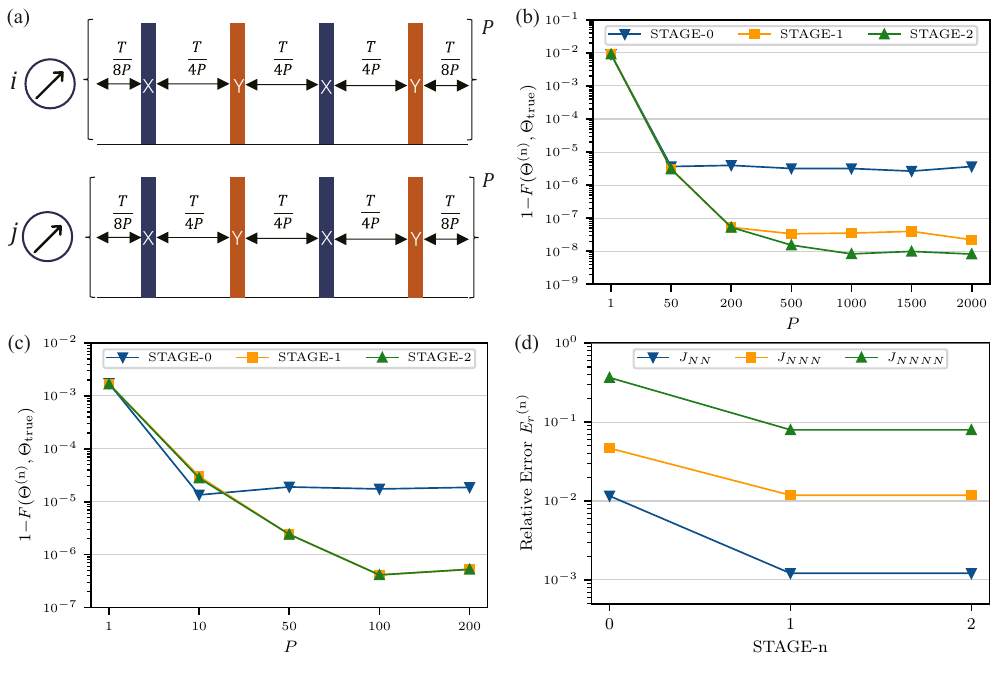} 
    \caption{(a) Schematic representation of periodic XY-4 pulse sequence applied to target qubits $i$ and $j$, which can be used to disentangle a many-body Hamiltonian into two-qubit subsystems. The X and Y denote the $\pi$ pulses along $x$ and $y$ directions, respectively. (b) Testing results for learning the Hamiltonian (\ref{eq: H3}) with $N=7$, displaying the infidelities $1-F(\bm{\Theta}^{(n)}, \bm{\Theta}_{\text{true}})$ for $n=0, 1, 2$ as a function of number of periods $P$ in the XY-4 sequence. (c) Testing results for learning the Hamiltonian (\ref{eq: H4}) with $N=7$. (d) Relative error $E_r^{(n)}$ ($n=0, 1$) for learning average coupling parameters $J_{\text{NN}}\in [-1, 1]$, $J_{\text{NNN}}\in [-0.1, 0.1]$, and $J_{\text{NNNN}}\in [-0.01, 0.01]$ with different magnitudes, corresponding to the results in (c). }
\label{fig: DDandTesting}
\end{figure*}

We now apply the two-qubit learning model to estimate the Hamiltonian parameters of more complex systems.
For larger systems, quantum dynamical decoupling (DD) methods can be employed to effectively disentangle a many-body system into two-qubit subsystems, making them suitable for our model~\cite{wang_hamiltonian_2015}. We adopt a periodic dynamical decoupling (PDD) pulse sequence, specifically the periodic XY-4, as illustrated in Fig.~\ref{fig: DDandTesting}(a). When the decoupling pulses act on the target qubits $i$ and $j$, the interaction between the qubits $i$ and $j$ is preserved while others are decoupled. In practical applications, they are applied to all other qubits except the target qubits to avoid operational errors. For a fixed total system evolution time  $T$, the decoupling effect is optimized by increasing the number of periods $P$ in the PDD sequence.

We test a generic Hamiltonian of Eq. (\ref{eq: Ham2qb1}) for an $N$-qubit system
\begin{equation}\label{eq: H3}
	H_3 = \sum_{i=1}^N \omega_i\sigma_z^i + \sum_{i\neq j = 1}^N J_{ij}\sigma_x^i\sigma_x^j,
\end{equation}
and further extend the model (\ref{eq: Ham2qb2}) to an even more complex scenario described by 
\begin{equation}\label{eq: H4}
	H_4 = \sum_{\langle i, j\rangle}J_{\text{NN}}^{ij\alpha\beta}\sigma_{\alpha}^i\sigma_{\beta}^j + \sum_{( i, j)}J_{\text{NNN}}^{ij\alpha\beta}\sigma_{\alpha}^i\sigma_{\beta}^j + \sum_{[ i, j]}J_{\text{NNNN}}^{ij\alpha\beta}\sigma_{\alpha}^i\sigma_{\beta}^j
\end{equation}
which involve nearest-neighbor (NN), next-nearest-neighbor (NNN), and next-next-nearest-neighbor (NNNN) interactions, with $\langle i, j\rangle$, $(i, j)$ and $[i, j]$ symbolizing these distinct connectivities.
For simplicity, we denote the average coupling constants of $J_{\text{NN}}^{ij\alpha\beta}$, $J_{\text{NNN}}^{ij\alpha\beta}$, and $J_{\text{NNNN}}^{ij\alpha\beta}$ as $J_{\text{NN}}$, $J_{\text{NNN}}$, and $J_{\text{NNNN}}$, respectively. We assume that $J_{\text{NN}}\in [-1, 1]$, $J_{\text{NNN}}\in [-0.1, 0.1]$, and $J_{\text{NNNN}}\in [-0.01, 0.01]$. 

The Hamiltonians (\ref{eq: H3}) and (\ref{eq: H4}) are decoupled into multiple two-qubit subsystems (\ref{eq: Ham2qb1}) and (\ref{eq: Ham2qb2}) through the PDD scheme of Fig. \ref{fig: DDandTesting}(a), respectively. Independent observations are performed for each subsystem, and the resulting data are fed into the pre-trained models in Figs. \ref{fig: lossfunction}(a) and \ref{fig: lossfunction}(b) to obtain the two-qubit Hamiltonian parameters. By iterating through the entire system, all parameters of the $N$-qubit Hamiltonians can be estimated. The fidelity of STAGE-$n$ between the whole estimated Hamiltonian parameters \(\bm{\Theta}^{(n)}\) and the true values \(\bm{\Theta}_{\text{true}}\) is evaluated using Eq. (\ref{eq: fidelity}).
Figures~\ref{fig: DDandTesting}(b)-\ref{fig: DDandTesting}(c) present the testing results of Hamiltonians (\ref{eq: H3}) and (\ref{eq: H4}) with $N=7$. As the number of periods $P$ in Fig. \ref{fig: DDandTesting}(a) increases, the fidelity at each stage improves significantly. This enhancement arises from the optimization of the decoupling effect, which reduces errors in the observational data provided to the two-qubit pre-trained models. Additionally, we observe that once the number of periods $P$ reaches a certain threshold, the fidelity tends to saturate, primarily due to the limitations of the pre-trained model at each stage. However, the precision of Hamiltonian parameter estimation can be further enhanced by utilizing a multi-stage learning model. As shown in Figs. \ref{fig: DDandTesting}(b) and \ref{fig: DDandTesting}(c), the fidelity of STAGE-1 improves two orders of magnitude compared to STAGE-0. In Fig. \ref{fig: DDandTesting}(b), the fidelity of STAGE-2 also shows an improvement of half an order of magnitude over STAGE-1, but no significant increase in Fig. \ref{fig: DDandTesting}(c).

We further investigate the sensitivity of the training model to parameters with different magnitudes (Appendix. \ref{seca: Sensitivity}). The motivation behind this setup is that parameters with smaller magnitudes contribute less to the loss function and fidelity, which may cause the model to fail to accurately learn these parameters even when the training process appears to converge. To quantify this effect, we use the relative error defined as:
\begin{equation}
E_r^{(n)} = \frac{\lvert \theta^{(n)} - \theta_{\text{true}} \rvert}{\lvert \theta_{\text{true}} \rvert},
\end{equation}
where $\theta^{(n)}$ is the predicted parameter after STAGE-$n$, and $\theta$ is the true value.
Figure~\ref{fig: DDandTesting}(d) shows that the relative error $E_r^{(0)}$ for parameter $J_{\text{NNNN}}$ remains close to $0.4$ in STAGE-0, indicating that the model fails to learn small parameters effectively by a single-stage neural network. In contrast, the $E_r^{(0)}$ for leading parameter $J_{\text{NN}}$ approaches the level of $10^{-2}$, in accordance with previous results. However, all $E_r^{(1)}$ of STAGE-1 improve one order of magnitude, demonstrating the effectiveness of the multi-stage approach in enhancing the learning precision of small-magnitude parameters. The $E_r^{(2)}$ without further improvement is limited by computation resources. 

\section{Robustness against noises}

\begin{figure}
    \centering
    \includegraphics[width=1\linewidth]{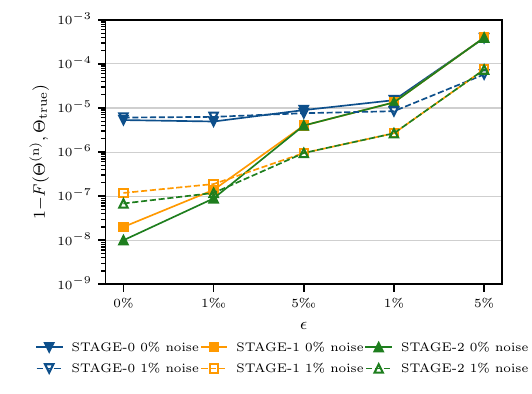}
    \caption{Robustness of the multi-stage Hamiltonian learning models (without and with 1\% training error) under varying intensity $\epsilon$ of validation noise. }
    \label{fig: robustness}
\end{figure}

In experiments, observed data inevitably suffers from environmental noise and readout errors. To validate the robustness of our multi-stage Hamiltonian learning model, we take the Hamiltonian (\ref{eq: Ham2qb1}) as an example and train models on datasets without and with a 1\% error. We then introduce varying levels of validation noise to these models. The results are illustrated in Fig. \ref{fig: robustness}. It can be found that the training capability of the multi-stage model becomes worse as the intensity of validation noise increases. Besides, the fidelity of STAGE-1 and STAGE-2 decreased more rapidly than that of STAGE-0, indicating that STAGE-0 is more robust, though the precision of Hamiltonian learning can be enhanced significantly by STAGE-1 and STAGE-2. However, the trained models at low noise levels maintain high performance, with both the training and validation infidelities remaining low. When the noise intensity reaches $1\%$, the multi-stage trained model without error begins to lose the capability of precision improvement. Additionally, while the maximum accuracy of the model is affected by the introduction of errors in the training data, we observed that as validation noise increases, the robustness of the model trained with errors surpasses that of the model trained without errors.

\section{Conclusions}

In conclusion, we demonstrate the feasibility and effectiveness of a multi-stage neural network framework for learning Hamiltonian parameters from single-qubit measurements. By employing a stage-by-stage training approach, our protocol achieves orders-of-magnitude improvements in parameter estimation accuracy compared to traditional and single-stage ML methods. Furthermore, our protocol shows the capability of parameter estimation of Hamiltonians in large systems by incorporating DD techniques and the robustness of the pre-trained models against noise.
Our work paves the way for scalable Hamiltonian learning with high precision, which is crucial for characterizing the dynamics of quantum systems, calibrating quantum devices, simulating the complex phenomena of condensed matter, and enhancing parameter sensing capabilities. 
Future directions include exploring the adaptability of our framework to open quantum systems, optimizing measurement strategies for higher-dimensional systems, and validating its performance on state-of-the-art quantum hardware. 

\section*{Acknowledgements}
This work was supported by the National Natural Science Foundation (Grant No. 11805008), Guangdong Basic and Applied Basic Research Foundation (Grant No. 2024A1515011406), Fundamental Research Funds for the Central Universities, Sun-Yat-Sen University (Grant No. 23qnpy63), Guangdong Provincial Key Laboratory (Grant No. 2019B121203005).

\appendix

\section{Optimization of sampling parameters}\label{seca: Optimization of sampling parameters}

\begin{figure}
    \centering
    \includegraphics[width=1\linewidth]{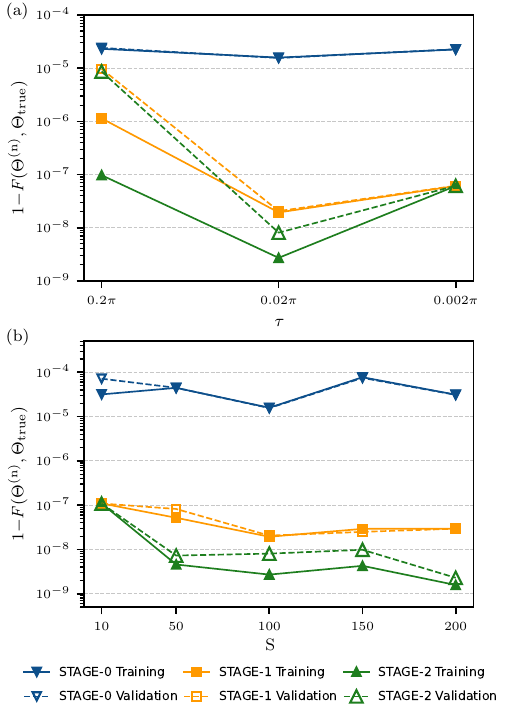}
    \caption{Performance of multi-stage Hamiltonian learning under varying sampling parameters. (a) Shows the infidelities, represented as $1-F(\bm{\Theta}^{(n)}, \bm{\Theta}_{\text{true}})$ ($n=0, 1, 2$), for training and validation across different sampling intervals $\tau$, with a fixed number of observations of $S=100$. (b) Displays the infidelities across different numbers of observations $S$, with a fixed sampling interval \( \tau = 0.02\pi \).}
    \label{fig: tauS}
\end{figure}

We investigated the effects of data sampling parameters, specifically the sampling interval \( \tau \) and the number of observations \( S \). Figure~\ref{fig: tauS}(a) examines the impact of the sampling interval \( \tau \) on the model performance. With a fixed number of observations \( S=100 \), the performance of STAGE-0 remains relatively stable regardless of the sampling interval. However, as multi-stage training progresses, the choice of sampling interval significantly influences the performance of STAGE-1 and STAGE-2. When the sampling interval is too small (\( 0.002\pi \)), the performance of STAGE-2 shows minimal improvement over STAGE-1, suggesting that excessively dense sampling does not yield additional useful information. Conversely, a sampling interval that is too large (\( 0.2\pi \)) improves training set performance but leaves validation fidelity unchanged, indicating limited generalization capability. Optimal results are achieved with a sampling interval of \( 0.02\pi \), leading to consistent improvements across all stages and ensuring robust validation set performance.
Figure~\ref{fig: tauS}(b) explores the influence of the number of observations \( S \) on model performance. With the sampling interval fixed at \( 0.02\pi \), the number of observations affects data characteristics in two significant ways: firstly, it determines the total data volume, which intuitively enhances model performance as more data is beneficial for training; secondly, it dictates the total evolution time of the system. Insufficient observations can result in a short evolution time, potentially missing the periodic dynamics of the system. While increasing the number of observations theoretically improves performance, results indicate that training performance stabilizes at \( S = 100 \). Beyond this point, further increases in \( S \) yield diminishing returns, especially when accounting for computational resource constraints.
Therefore, our analysis identifies \( \tau = 0.02\pi \) and \( S = 100 \) as the optimal parameter configuration. This choice balances model performance and computational resource consumption.

\section{Error statistics}\label{seca: Error distributions}

\begin{figure*}
    \centering
    \includegraphics[width=0.9\textwidth]{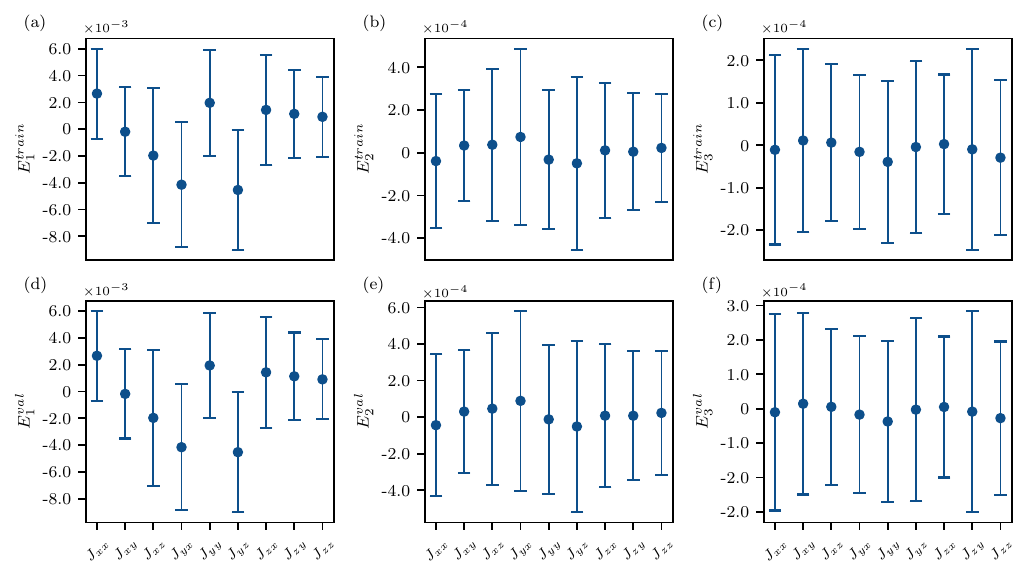} 
    \caption{Error statistics for learning the Hamiltonian $H_2$ with nine parameters $J_{\alpha\beta}$ for $\alpha, \beta \in\{x, y, z\}$. (a)-(c) show the training results for STAGE-0, STAGE-1, and STAGE-2, respectively. (d)-(f) illustrate the validation results. The solid points represent the mean values of errors across 300,000 datasets, and the error bars indicate the standard deviation of these errors.}
    \label{fig: errordistrubitionH2}
\end{figure*}

Figures~\ref{fig: errordistributionH1} and \ref{fig: errordistrubitionH2} present the error statistics of the parameters of Hamiltonian $H_1$ and $H_2$ calculated from the error distributions at each stage for both the training and validation datasets for STAGE-0, STAGE-1, and STAGE-2. As the training stage progresses, the error gradually decreases, and the standard deviation of error also becomes smaller orders of magnitude. The multi-stage Hamiltonian learning model effectively reduces both training and validation errors through iterative error learning. Figures \ref{fig: errordistributionH1}(c) and \ref{fig: errordistributionH1}(f) demonstrate that even at STAGE-2, the errors continue to decrease, reflecting the potential performance and generalization capability of multi-stage neural networks.

\begin{figure*}
    \centering
    \includegraphics[width=0.9\textwidth]{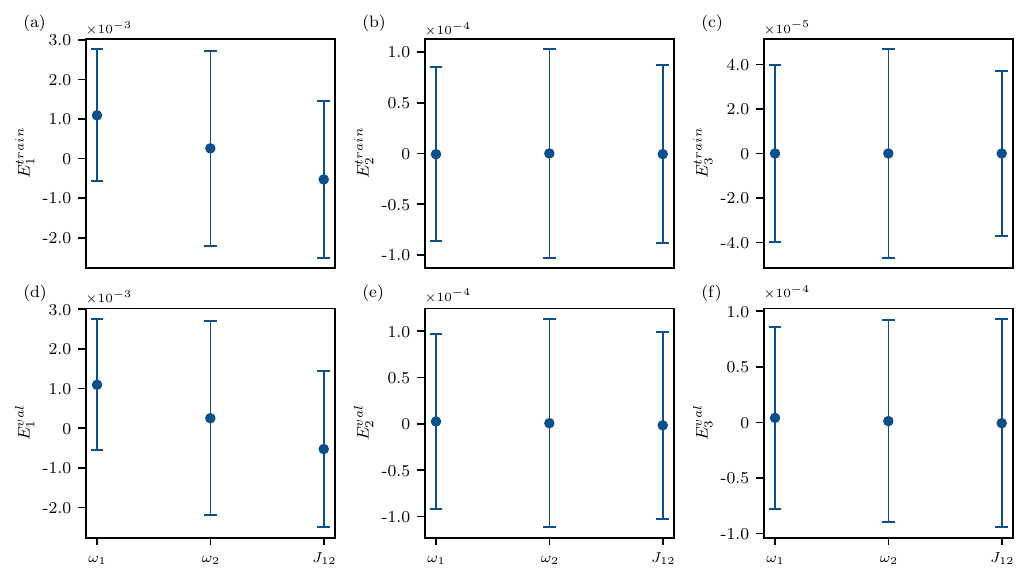} 
    \caption{Error statistics for learning the Hamiltonian $H_1$ with three parameters $\omega_1$, $\omega_2$, and $J_{12}$. (a)-(c) show the training results for STAGE-0, STAGE-1, and STAGE-2, respectively. (d)-(f) illustrate the validation results. The solid points represent the mean values of errors across 300,000 datasets, and the error bars indicate the standard deviation of these errors.}
    \label{fig: errordistrubitionH1}
\end{figure*}

%The error distributions are visualized using box plots, a tool from descriptive statistics. In each box plot, the central line represents the median error, the box edges mark the first and third quartiles (Q1 and Q3), and the whiskers extend to 1.5 times the interquartile range (IQR). Outliers beyond this range are depicted as individual points.

\section{Data correlations}\label{seca: Data correlations}
In statistics, the Pearson correlation coefficient (PCC) \( r \) quantifies the linear correlation between two variables \( X \) and \( Y \), with values ranging from \( -1 \) to \( 1 \)~\cite{benesty2009noise}. It is defined as
\begin{equation}
r = \frac{\sum_{i=1}^n (x_i - \bar{x})(y_i - \bar{y})}{\sqrt{\sum_{i=1}^n (x_i - \bar{x})^2} \cdot \sqrt{\sum_{i=1}^n (y_i - \bar{y})^2}}.
\end{equation}
In information theory, mutual information (MI) quantifies the dependency between two variables, reflecting the amount of information one variable provides about another~\cite{PhysRevE.69.066138}. Unlike the PCC, MI captures both linear and nonlinear relationships and is defined as:
\begin{equation}
I(X; Y) = \sum_{x \in X} \sum_{y \in Y} p(x, y) \log \frac{p(x, y)}{p(x)p(y)}.
\end{equation}

\begin{figure}
    \centering
    \includegraphics[width=1\linewidth]{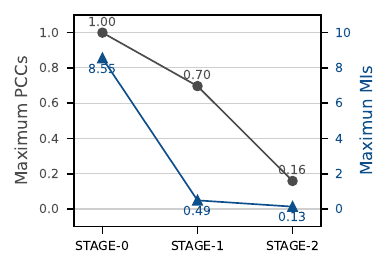}
    \caption{Maximum Pearson correlation coefficient (PCC) and maximum mutual information (MI) between input data and target outputs at each stage. The PCC reflects linear correlations, while the MI captures both linear and nonlinear dependencies. The results indicate a progressive decrease in correlation as the stages advance, suggesting a diminishing relevance of input data to the residual errors.}
  \label{fig: PCCMI}
\end{figure}

From the results presented in Fig.~\ref{fig: PCCMI}, we observe that starting from STAGE-0, the input data and target outputs exhibit significant linear and nonlinear correlations, evidenced by high values of PCC and MI. This provides substantial relational information for effective model training. However, as training progresses to STAGE-2, both the maximum PCC and MI values gradually decrease, approaching zero. This decline indicates a significant weakening of the correlation between the input data and target outputs in STAGE-2, making it challenging for the model to extract meaningful learning signals from the residual error information. Consequently, this phenomenon explains why increasing model complexity does not yield improvements in STAGE-2 training performance. These findings offer valuable insights for future enhancements in training data generation strategies and model architecture optimization.

\section{Sensitivity to small parameters}\label{seca: Sensitivity}

\begin{figure}
    \includegraphics[width=0.48\textwidth]{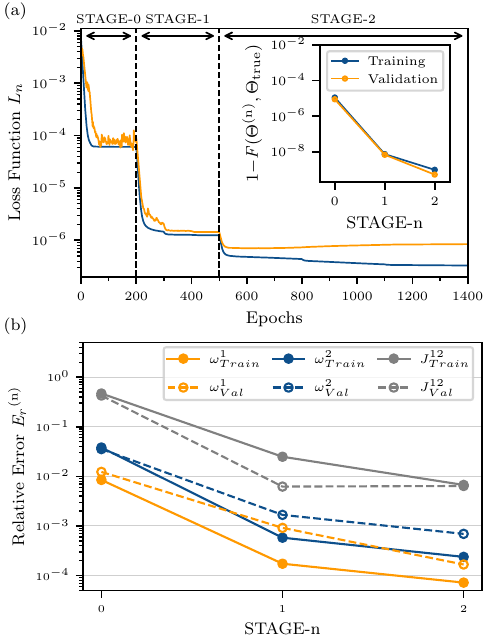} 
    \caption{Results of multi-stage learning for a 2-qubit Hamiltonian with parameters of different magnitudes: $\omega_1 \in [-1, 1]$, $\omega_2 \in [-0.1, 0.1]$, $J_{12} \in [-0.01, 0.01]$. (a) Loss function $L_n$ as a function of epochs. The inset shows the infidelity for each stage. (b) Relative error $E_r^{n}$ for $n=0,1,2$, showing the successive enhancement of parameter estimation accuracy through the multi-stage Hamiltonian learning.
    }
    \label{fig: sensitivity}
\end{figure}

We investigate the training sensitivity of another model to parameters of different magnitudes. We consider the Hamiltonian defined in Eq.~\ref{eq: Ham2qb1}, where the parameter ranges were set to $\omega_1 \in [-1, 1]$, $\omega_2 \in [-0.1, 0.1]$, and $J_{12} \in [-0.01, 0.01]$. Figure~\ref{fig: sensitivity}(a) shows the convergence of the loss function, similar to the results of Fig. \ref{fig: lossfunction}(a). Even in the presence of small parameters, the multi-stage Hamiltonian learning progressively refines its predictions by learning residual errors, leading to significant improvements in training accuracy and fidelity. The inset of Fig.~\ref{fig: sensitivity}(a) clearly shows the training performance at each stage, validating the effectiveness of our proposed multi-stage training strategy. 

However, parameters with smaller magnitudes contribute less to the loss function and overall fidelity, potentially leading to the model's inability to accurately learn these parameters, even when the training process appears to converge.  As shown in Fig.~\ref{fig: sensitivity}(b), the relatively high value of $E_r^{(0)}$ in STAGE-0 for $J_{12}$ indicates that a single-stage neural network struggles to effectively learn small parameters. But the $E_r^{(2)}$ for $J_{12}$ achieves a lower value than $E_r^{(0)}$ for $\omega_1$ with the multi-stage neural network, demonstrating that we can stage-by-stage enhance the learning precision of small-magnitude parameters.

\bibliographystyle{apsrev4-1}
%\bibliography{Ref_MSHL}


\end{document}